# All-in-one foundational models learning across quantum chemical levels

Yuxinxin Chen[1] and Pavlo O. Dral[1,2*]

[1]*State Key Laboratory of Physical Chemistry of Solid Surfaces, Fujian Provincial Key Laboratory of Theoretical and Computational Chemistry, Department of Chemistry, and College of Chemistry and Chemical Engineering, Xiamen University, Xiamen 361005, China.*

[2]*Institute of Physics, Faculty of Physics, Astronomy, and Informatics, Nicolaus Copernicus University in Toruń, ul. Grudziądzka 5, 87-100 Toruń, Poland.*

*Email: dral@xmu.edu.cn*

## Abstract

Machine learning (ML) potentials typically target a single quantum chemical (QC) level while the ML models developed for multi-fidelity learning have not been shown to provide scalable solutions for foundational models. Here we introduce the all-in-one (AIO) ANI model architecture based on multimodal learning which can learn an arbitrary number of QC levels. Our all-in-one learning approach offers a more general and easier-to-use alternative to transfer learning. We use it to train the AIO-ANI-UIP foundational model with the generalization capability comparable to semi-empirical GFN2-xTB and DFT with a double-zeta basis set for organic molecules. We show that the AIO-ANI model can learn across different QC levels ranging from semi-empirical to density functional theory to coupled cluster. We also use AIO models to design the foundational model Δ-AIO-ANI based on Δ-learning with increased accuracy and robustness compared to AIO-ANI-UIP. The code and the foundational models are available at https://github.com/dralgroup/aio-ani; they will be integrated into the universal and updatable AI-enhanced QM (UAIQM) library and made available in the MLatom package so that they can be used online at the XACS cloud computing platform (see https://github.com/dralgroup/mlatom for updates).

**Introduction**

The rise of machine learning interatomic potentials (MLIPs) has a transformative effect on quantum chemistry (QC).[1-13] The increasing availability of QC data[14] leads to the current trend of training universal ML interatomic potentials (UIPs) transferable across different chemical systems signaling the watershed moment in QC.[15-18] In any case, the choice of the reference QC level greatly influences how much data can be generated and what accuracy can be achieved: the more accurate the level, the more expensive it is and less data can be generated. Also, no single level of theory is suitable for all possible applications. As the field quickly develops, it becomes apparent that the multitude of available QC levels presents the major source of opportunities and challenges in creating MLIPs and UIPs.

Over the years, several approaches were suggested to exploit different QC levels which typically have to deal with the more data available at the more approximate levels and less data at more accurate ones.[19] One of the powerful approaches is transfer learning (TL) where the model is first trained on one more abundant less accurate level and then fine-tuned to the fewer more accurate data.[20] In the end, in TL we end up with having two separate models targeting either approximate or more accurate QC levels. A different approach is Δ-learning[21] and related approaches[22, 23] where only one model is trained to correct the low-level, baseline QC method to the target higher-level QC method but it is restricted to the specific combination of the baseline and target levels and requires the evaluation at the QC level at the inference time. Another approach is co-kriging[24] where the models are trained on several data simultaneously, but due to the bad scaling with the increasing number of the training points this approach did not find a wide-spread use and was never used for UIP. Other related approaches are including the predictions of the lower-level method in the features,[25] multi-task learning,[26] hierarchical machine learning,[27] multilevel combination technique,[28] multilevel learning,[29] and different variants of multi-fidelity learning[30].

Despite all the attempts, the use of the multitude of different levels is underutilized. One of the major obstacles is the absence of scalable and easily extendable model architecture which would allow training on big data and an arbitrary number of QC methods and, if required, easily adjustable to more levels. In this work, we propose an all-in-one (AIO) model architecture based on the idea of multimodal learning, where we simply include the reference level of theory as an input feature alongside the geometric features. We show that AIO model architecture provides a powerful tool for learning on multi-level data: a single model can make predictions at any of the reference levels eliminating the need to train separate MLIP for each level. Based on this architecture we create a series of UIPs that are transferable across the chemical compound space and can give predictions at levels ranging from semi-empirical to density functional theory (DFT) to CCSD(T)/CBS (complete basis set). In addition, we show that they provide a straightforward way to create Δ-learning corrections for a large number of combinations of the baseline and target QC levels resulting in AI-enhanced quantum mechanical methods with greatly improved robustness and accuracy compared to the corresponding UIPs without the QC baseline. Our AOI universal models (both UIPs and Δ-learning-based universal models) serve as foundational models for learning across the QC levels. They will be made available to the community via MLatom[31] and integrated into our universal and updatable AI-enhanced QM (UAIQM)[32] library which is available for computations online at https://XACScloud.com.

**Results and discussion**

*Model architecture*

All-in-one (AIO) machine learning interatomic potential needs to learn energies based on multimodal information: geometries and level of theory. In this work, we have built the required AIO architecture based on the ANI-type modification[15] of the Behler–Parrinello neural networks[33] (NNs) which encode the geometric information via ANI-type atomic environment vectors (AEV) used as features (Figure 1). The networks' activation function and cutoffs were modified in the same way as in the AIQM1[34] and the first generation of the UAIQM[32] methods. ANI-AEV features are generated for each atom based on the element type. To featurize the QC level, we use the one-hot encoding and append this additional feature to the ANI-AEV. These concatenated features are passed to the networks corresponding to each element type. The outputs of networks are the atomic energies which are summed up to the total energy at the QC level to be learned/predicted. We call the resulting network architecture AIO-ANI.

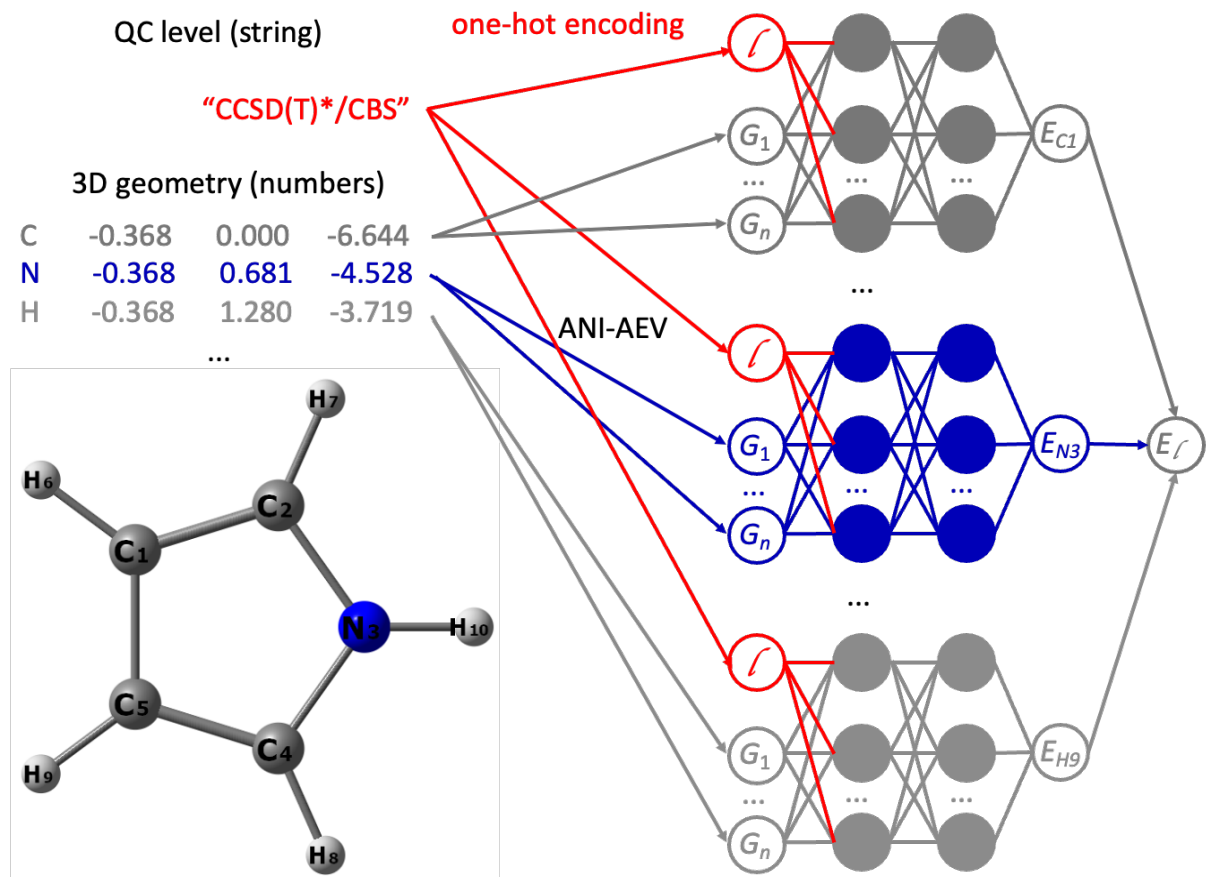

**Figure 1.** All-in-one ANI-type neural network architecture for learning across QC levels. The model is based on ANI-type atomic environment vectors (AEV) encoding the geometric information and adds information about the level of theory through additional input feature obtained via one-hot encoding from the string. These features are generated for each atom based on element type and the outputs of networks are the atomic energies which are summed up to the total energy at the required level.

This choice of the network has the benefit of being based on one of the most time-tested MLIP architectures successfully utilized in numerous UIPs (ANI-1,[35] ANI-1x,[36] ANI-1ccx,[37] ANI-2x[38]) and AI-enhanced QM methods (AIQM1[34], UAIQM[32]). Compared to the models based on equivariant networks ANI MLIPs are less data-efficient but their high computational speed of training and inference make up for this disadvantage.[39] In addition, there are available multi-level QC data sets[40, 41] specifically curated for ANI-based UIPs which makes ANI-based architecture one of the most optimal choices. It is also increasingly emphasized by experts that hunting for more data-efficient solutions might be not the most practical approach as it might be easier to deal with simpler architectures and generate more data.[42]

We perform the training of the AIO-ANI networks on the modified ANI-1ccx [40] data set which contains ca. 4.5 M energies and forces at the $\omega$B97X[43]/def2-TZVPP[44] level and ca. 0.5 M energies at the CCSD(T)*/CBS level (a special extrapolation scheme to achieve CCSD(T)/CBS level). The data set comprises off-equilibrium structures of small molecules and was used to train aforementioned UIPs ANI-1x and ANI-1ccx. We modify it by excluding the explicit D4 correction[45] (for the $\omega$B97X functional) from the CCSD(T)/CBS level and adding energies and forces calculated for the ca. 4.5 M configurations at the semi-empirical QC levels GFN2-xTB* and ODM2* (the corresponding methods GFN2-xTB[46] and ODM2[47] without explicit dispersion corrections). While using the AIO-ANI networks for inference, we add the explicit D4 corrections back. This is done analogously to the handling of the explicit dispersion corrections in AIQM1[34] and the first generation of UAIQM methods[32] because the ANI network is local and it might be beneficial to treat the long-range dispersion contributions explicitly. Before training, the data for each QC level is centered as usual via calculating the self-atomic energies ($E_{\mathrm{SAE}}$). They are added back during the inference.

The beauty of the AIO-ANI architecture is that once the model is trained, the inference can be done for the target QC level $l$ by simply providing the level as a string in input to the NN function $f_{\mathrm{AIO\text{-}ANI\text{-}NN}}$ (along with the geometry $\boldsymbol{R}$):

$$E_{\mathrm{AIO\text{-}ANI}}(\boldsymbol{R}, l) = f_{\mathrm{AIO\text{-}ANI\text{-}NN}}(\boldsymbol{R}, l) + E_{\mathrm{SAE}}(\boldsymbol{R}, l) + E_{\mathrm{D4}}(\boldsymbol{R}). \qquad (1)$$

This provides a very flexible solution to calculating $\Delta$-learning correction for any pair of the QC levels on which the network was trained. This correction can be then added to the actual baseline QC predictions (with removed dispersion corrections) yielding an easy way to construct a multitude of $\Delta$-learning-based AI-enhanced QM methods similar to AIQM1 and UAIQM without the need to train separate networks for each combination of the baseline and target levels. The predictions with $\Delta$-learning-based models are:

$$E(\boldsymbol{R}, l) = E_{\mathrm{baseline}^*}(\boldsymbol{R}) + E_{\mathrm{AIO\text{-}ANI}}(\boldsymbol{R}, l_{\mathrm{target}}) - E_{\mathrm{AIO\text{-}ANI}}(\boldsymbol{R}, l_{\mathrm{baseline}}) + E_{\mathrm{D4}}(\boldsymbol{R}). \qquad (2)$$

**AIO-ANI foundational model**

We train the above AIO-ANI architecture on the two QC levels: DFT (ωB97X/def2-TZVPP) and the gold-standard coupled cluster (CC, namely CCSD(T)/CBS). We use the ANI-1ccx data set with ca. 4.5 M conformations with DFT energies and forces and 0.5 M energies at the CC level. The resulting AIO-ANI-UIP is a single model that can be used to give predictions at both DFT and CC levels with the high speed of the standard ANI MLIP. This model has an accuracy approaching both levels for the data within the distribution of the ANI-1ccx data set: the validation errors are ca. 1.2 kcal/mol for each level. Of course, the more stringent test is the generalization ability as judged on independent benchmark sets. Hence, we analyze the errors on the standard GMTKN55[48] benchmark set to put AIO-ANI-UIP performance in perspective, by comparing it to several QC approaches (Figure 2). The weighted mean absolute deviation-2 (WTMAD-2)[48] of AIO-ANI-UIP is smaller when making predictions at the CC level reflecting the higher accuracy of the reference level, although the data set is much smaller compared to the DFT-level data. This indicates that the model can effectively capture the underlying correlations between the levels and learn the higher accuracy of CC while benefitting from the better coverage of chemical space with DFT.

Our AIO-ANI-UIP model accuracy is overall comparable to both the popular semi-empirical GFN2-xTB and DFT B3LYP-D4/6-31G* methods while the speed is much higher (Figure 2, note that these results are for the CHNO-containing closed-shell neutral species).

| | Transfer Learning (CC) | AIO (DFT) | AIO (CC) | Δ-AIO | GFN2-xTB | B3LYP-D4/ 6-31G* |
|---|---|---|---|---|---|---|
| **basic properties and reaction energies for small systems** | | | | | | |
| G2RC | 69.28 | 15.94 | 16.41 | 18.79 | 29.52 | 10.66 |
| FH51 | 19.37 | 5.34 | 4.01 | 5.40 | 11.21 | 5.75 |
| TAUT15 | 1.12 | 2.18 | 1.61 | 1.07 | 0.99 | 1.59 |
| DC13 | 42.34 | 110.71 | 109.46 | 6.43 | 26.61 | 5.99 |
| WTMAD-2 | 17.97 | 14.84 | 13.56 | 5.38 | 10.19 | 5.50 |
| **reaction energies for large systems and isomerisation reactions** | | | | | | |
| DARC | 1.64 | 2.86 | 1.99 | 1.62 | 17.44 | 2.08 |
| BSR36 | 3.18 | 5.22 | 1.73 | 5.39 | 2.76 | 3.14 |
| CDIE20 | 2.12 | 3.05 | 2.86 | 1.11 | 1.80 | 1.95 |
| ISO34 | 1.43 | 1.33 | 0.88 | 1.51 | 6.90 | 2.89 |
| ISOL24 | 5.44 | 6.20 | 5.47 | 3.54 | 14.59 | 3.65 |
| C60ISO | 54.34 | 73.36 | 75.20 | 16.25 | 5.80 | 2.55 |
| WTMAD-2 | 5.37 | 7.47 | 5.60 | 4.58 | 8.64 | 4.83 |
| **reaction barrier heights** | | | | | | |
| BHPERI | 17.41 | 18.75 | 15.40 | 3.14 | 9.49 | 4.02 |
| BHDIV10 | 13.75 | 10.10 | 9.91 | 4.20 | 10.83 | 4.22 |
| INV24 | 5.76 | 8.38 | 7.91 | 2.29 | 2.63 | 1.68 |
| BHROT27 | 1.31 | 1.70 | 1.54 | 0.35 | 0.91 | 0.75 |
| PX13 | 8.19 | 9.37 | 9.90 | 2.11 | 3.65 | 8.34 |
| WCPT18 | 7.28 | 8.40 | 9.46 | 1.81 | 3.83 | 6.49 |
| WTMAD-2 | 9.34 | 10.69 | 9.50 | 2.14 | 5.25 | 3.53 |
| **intermolecular noncovalent interactions** | | | | | | |
| ADIM6 | 2.72 | 0.99 | 1.33 | 0.61 | 1.15 | 0.47 |
| S22 | 5.18 | 5.79 | 5.91 | 1.64 | 0.76 | 2.58 |
| S66 | 2.97 | 2.78 | 2.70 | 0.85 | 0.73 | 2.10 |
| WATER27 | 1.57 | 18.91 | 0.94 | 17.80 | 1.50 | 55.36 |
| WTMAD-2 | 12.09 | 11.98 | 11.30 | 4.16 | 2.86 | 9.34 |
| **intramolecular noncovalent interactions** | | | | | | |
| IDISP | 33.51 | 11.73 | 9.27 | 6.89 | 6.78 | 1.82 |
| ACONF | 0.70 | 0.32 | 0.21 | 0.15 | 0.19 | 0.19 |
| Amino20x4 | 1.02 | 1.30 | 1.44 | 0.34 | 0.98 | 0.71 |
| PCONF21 | 2.54 | 2.75 | 2.31 | 1.76 | 1.76 | 1.88 |
| MCONF | 1.33 | 2.00 | 1.67 | 1.57 | 1.72 | 1.37 |
| SCONF | 1.72 | 3.03 | 3.11 | 0.71 | 1.64 | 3.66 |
| BUT14DIOL | 0.57 | 0.60 | 0.34 | 0.28 | 1.25 | 2.44 |
| WTMAD-2 | 10.56 | 11.56 | 10.41 | 5.59 | 9.97 | 12.15 |
| total WTMAD-2 | 10.54 | 11.10 | 9.87 | 4.69 | 7.93 | 8.31 |

**Figure 2. Performance of AIO models trained with CCSD(T)*/CBS and ωB97X/def2-TZVPP level data as benchmarked on GMTKN55 (CHNO subset with closed-shell neutral species).** AIO (DFT) denotes predictions by the AIO-ANI-UIP model targeting the DFT level and AIO (CC) – targeting the coupled cluster level. The transfer learning model was first pre-trained on the same DFT-level and then fine-tuned on the same coupled cluster-level data. Δ-AIO-ANI model is adding the difference between AIO (CC) and AIO (DFT) predictions to the DFT baseline. All ML models include explicit dispersion corrections.

**All-in-one, multimodal learning is a better alternative to transfer learning**

An obvious alternative to AIO learning approach is transfer learning (TL). The advantages of AIO over transfer learning are: 1) AIO is capable of making predictions at any level it saw during training while when using transfer learning, we end up with two separate models for each level, 2) TL is typically used for two levels, while AIO can be applied to an arbitrary number of levels, 3) TL is a two-step process (pre-training and fine-tuning) and requires the choice of how the parameters are fine-tuned (e.g., what layers to freeze, how to change the learning rate, etc.), while AIO training is done in one step. Here we analyze how AIO compares to TL for the same task of creating the UIP on two levels (DFT and CC) as above. AIO training converges much faster even than the fine-tuning step of TL (in 1000 epochs vs 1750 epochs, Figure 3), while the pre-training step in TL also takes lots of effort (ca. 2000 epochs). In the end, AIO-UIP is slightly more accurate (WTMAD-2 is 9.87 kcal/mol) than the model produced by TL (WTMAD-2 of 10.54 kcal/mol, Figure 2). Overall, the AIO-UIP model seems to have more consistent performance across different type of properties.

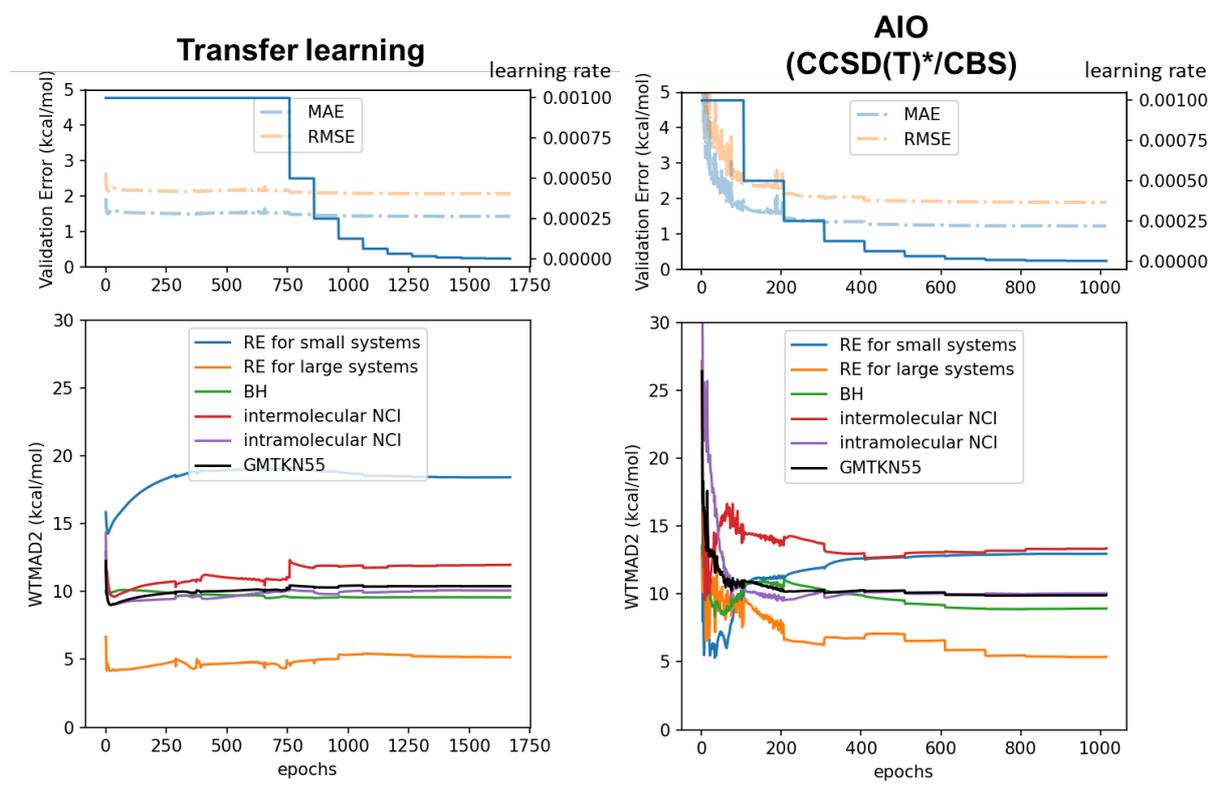

**Figure 3. Comparison of transfer learning and all-in-one learning.** The tested models are trained on CCSD(T)*/CBS and ωB97X/def2-TZVPP-level data. Both are using S30L in the validation set to stabilize the training process (see text). The training is terminated when the learning rate limit ($10^{-6}$) is reached.

**Improving all-in-one, multimodal learning with Δ-learning**

In all the discussion about different approaches for learning multiple QC levels, we should not forget that they are not necessarily rivals but can often be used synergetically.[49] One way to improve the stability of ML predictions is to use ML only for correcting the baseline QC level with the Δ-learning approach, which typically imparts greater robustness and accuracy. Here, we explore this possibility by using the AIO model to generate the Δ-learning correction. Since our AIO-ANI-UIP was trained on DFT and CC, we can easily generate the CC−DFT correction with the AIO-ANI-UIP and add it to the actual DFT predictions per Eq. 2. The resulting Δ-AIO-ANI foundational model overall performs significantly better than the pure ML alternative. Δ-AIO-ANI has the WTMAD-2 of 4.69 kcal/mol – only half of the WTMAD-2 of AIO-ANI-UIP and significantly below the GFN2-xTB or B3LYP/6-31G* (Figure 2).

**Towards learning from more heterogenous data and more levels**

As a proof-of-concept, we train the AIP-ANI model on more heterogeneous data and more levels. If we have only a few levels (CC and DFT with different basis sets), the validation errors drop quite fast but for more levels (e.g., including semi-empirical GFN2-xTB and ODM2), the validation error keeps dropping with many more epochs (Figure 4). More interesting is an analysis of how the generalization error evaluated on the GMTKN55 set changes with different compositions of data and the number of epochs. We see that, e.g., including forces at the lower QC levels can potentially help to achieve smaller generalization errors.

However, more research is required while generating UIP models as follows from our analysis below. The generalization error has an erratic dependence on the number of epochs – an important observation that was not reported before as far as we are aware. This has to be paid attention to in the development of UIPs. The erratic dependence indicates that it is easy to overfit the model and that monitoring error at the validation set taken from the same distribution as the training set, does not help to avoid overfitting. Another problem is that the generalization errors from predictions targeting CC level are not necessarily better than targeting lower-level CC. These problems might be related. We solve them by adding an external validation set not used in training: the S30L data set[50] because we found that the errors for the noncovalent interactions increase the most during training (Figure 5). This helped to stabilize the generalization error overall (Figure 5 and Figure 3) and allowed us to obtain the AIO-ANI-UIP and Δ-AIO-ANI models described above (Figure 2).

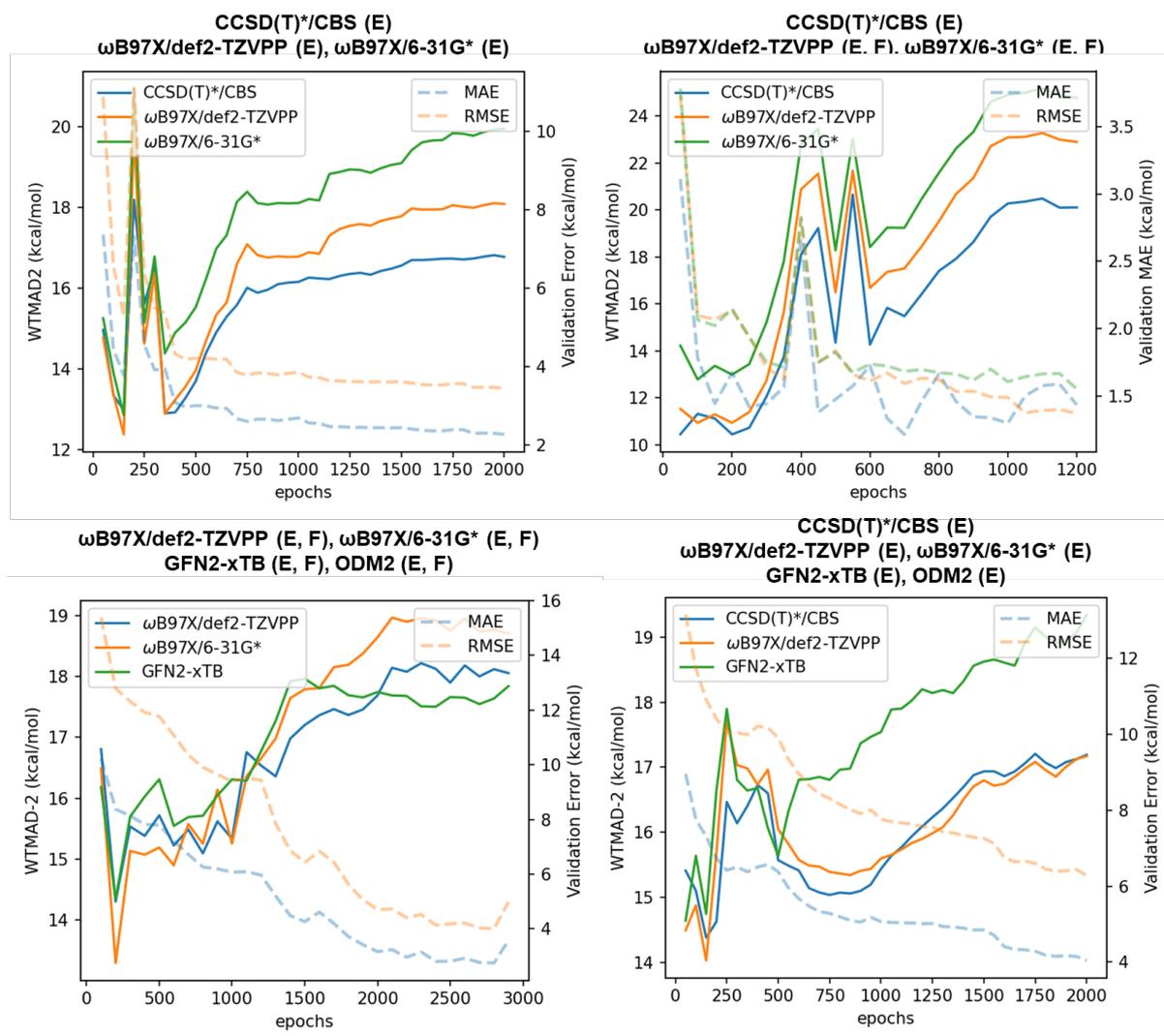

**Figure 4. Validation and generalization error of AIO models as the training progresses on different level combinations of data.** The title for each subplot indicates the levels of data used in training. E and F in the parentheses indicate whether energies or forces are used.

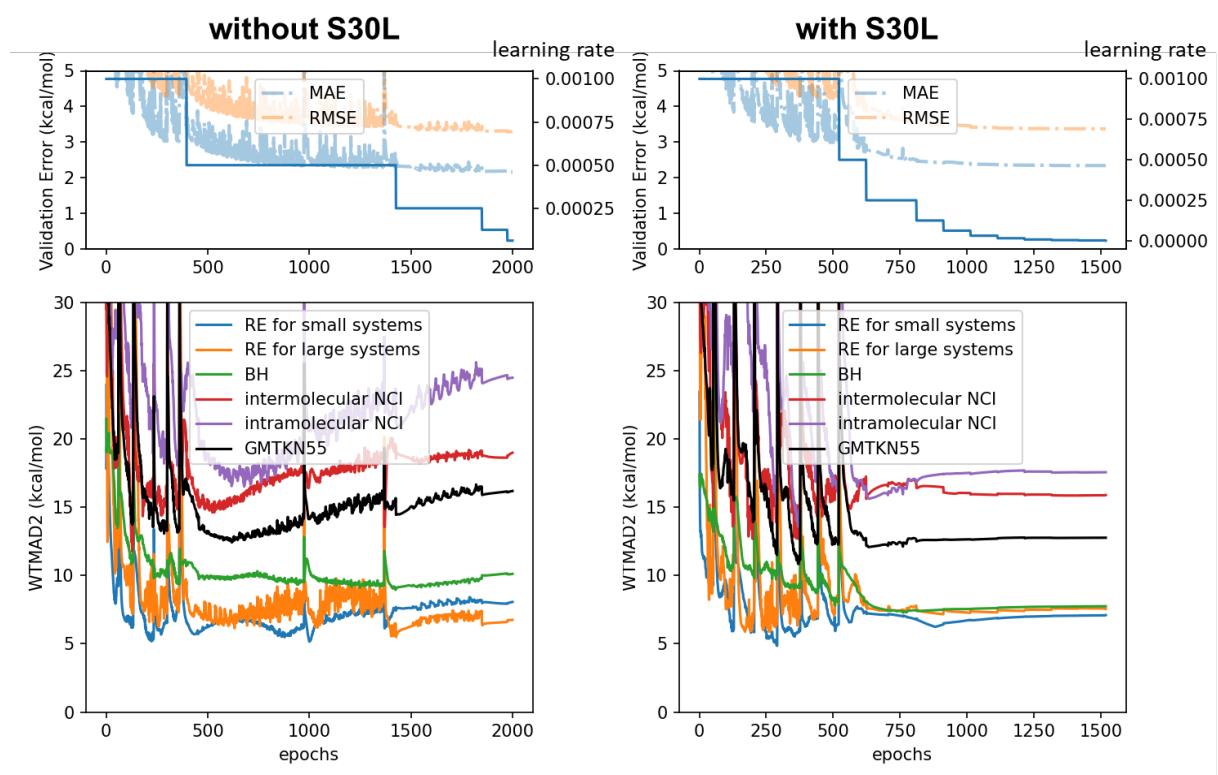

**Figure 5. Comparison of AIO model trained only on the CCSD(T)\*/CBS-level data with and without S30L as an external validation set.** 18 relative energies of S30L are used as the external validation set. The ratio between the errors in energies for the internal validation set (part of the ANI-1ccx data set) and S30L is 1:1. The training is terminated when the learning rate limit ($10^{-6}$) is reached.

Despite the problems with generalization stability, the results in Figure 4 indicate that it is in principle possible to train the AIO-ANI model on many levels as the models trained with three, four, and five levels could achieve decent generalization performance in at least some epochs. Such models can readily be used for constructing many Δ-learning-based models for different combinations of the baseline and target levels without the need to train separate models for each combination.

**Conclusions**

In this work, we presented a novel, all-in-one, approach for learning across different quantum chemical levels in one model. This approach is capable of learning from the heterogenous data in a single step, is scalable to big data and arbitrary number of QC levels. We used this AIO approach to train the foundational AIO-ANI-UIP model which has a performance close to the popular semi-empirical and DFT methods but with the cost of fast ML interatomic potential. This single model can make estimates of energies and forces targeting different QC levels.

We also showed that the AIO approach provides an attractive alternative to transfer learning due to its simplicity and generalizability. It can readily be integrated into Δ-learning models as the AIO model can generate corrections for different combinations of the baseline and target level. We exploited this to create the Δ-AIO-ANI model using the DFT method as a baseline and having better accuracy than, e.g., AIO-ANI-UIP and B3LYP/6-31G*.

**Computational details**

The D4 dispersion corrections were calculated with the dftd4 program.[45, 51] AIO-ANI models were built based on TorchANI[15]. The xtb program[52] was used for the GFN2-xTB calculations. ODM2 contributions were calculated with the MNDO program[53]. PySCF[54-56] was used for the DFT calculations.

**Data availability**

The calculations are based on the publicly available ANI-1ccx data set.

**Code availability**

The code and the foundational models are available at https://github.com/dralgroup/aio-ani; they will be made available in MLatom (see https://github.com/dralgroup/mlatom for updates).

**Author contributions**

Y.C.: implementations and calculations. P.O.D: conceptualization and supervision. Both authors analyzed the results and wrote the manuscript.

**Acknowledgments**

P.O.D. acknowledges funding from the National Natural Science Foundation of China (funding via the Outstanding Youth Scholars (Overseas, 2021) project) and via the Lab project of the State Key Laboratory of Physical Chemistry of Solid Surfaces.